\documentclass[aps,prl,twocolumn,superscriptaddress,appendix,showpacs,floatfix,nobibnotes]{revtex4}
\usepackage{epsfig}
\usepackage{epstopdf}

\usepackage{graphicx}
\usepackage{longtable}
\usepackage{CJK}
\usepackage{color}
\usepackage{mathptmx, courier, pifont}
\usepackage[scaled=0.92]{helvet}
\usepackage[T1]{fontenc}
\usepackage{textcomp}

\begin{document}

\begin{CJK*}{GB}{}

\title{B(E2) anomaly and triaxial deformation within a two-fluid  SU(3) symmetry}

\author{Wei Teng}
\affiliation{Department of Physics, Liaoning Normal University,
Dalian 116029, P. R. China}

\author{Sheng Nan Wang}
\affiliation{Department of Physics, Liaoning Normal University,
Dalian 116029, P. R. China}

\author{Yu Zhang }\email{dlzhangyu_physics@163.com}
\affiliation{Department of Physics, Liaoning Normal University,
Dalian 116029, P. R. China}

\author{Xian Zhi Zhao}
\affiliation{Department of Physics, Liaoning Normal University,
Dalian 116029, P. R. China}

\author{Xi Deng}
\affiliation{Department of Physics, Liaoning Normal University,
Dalian 116029, P. R. China}

\author{Xiao Tong Li}
\affiliation{Department of Physics, Liaoning Normal University,
Dalian 116029, P. R. China}

\begin{abstract}
The correlation between $B(E2)$ structure and triaxial deformation has been investigated within the framework of the proton-neutron boson model. The analysis reveals that the distinctive feature, characterized by $B(E2;4_1^+\rightarrow2_1^+)/B(E2;2_1^+\rightarrow0_1^+)<1.0$ along with $E(4_1^+)/E(2_1^+)>2.0$, can emerge from the triaxial SU(3) symmetry inherent in two-fluid boson systems, attributed to band-mixing effects. This suggests a symmetry-based understanding of the anomalous $E2$ transitions observed in experiments.
\end{abstract}
\pacs{21.60.Fw, 21.60Ev, 21.10Re}

\maketitle

\end{CJK*}

\begin{center}
\vskip.2cm\textbf{I. Introduction}
\end{center}\vskip.2cm

The emergence of collective features is one of the most important and striking characteristics of complex nuclear many-body systems.
The associated collective modes can be effectively demonstrated within the framework of the interacting boson model (IBM) using group or algebraic language~\cite{Iachellobook}. Notable examples include
U(5) (spherical vibrator), SU(3) (axially-deformed rotor), and O(6) ($\gamma$-unstable rotor). A common feature of these conventional modes is $B_{4/2}\equiv B(E2;4_1^+\rightarrow2_1^+)/B(E2;2_1^+\rightarrow0_1^+)>1.0$, along with $R_{4/2}=E(4_1^+)/E(2_1^+)\geq2$, which are consistent with various theoretical calculations and extensive experimental data
on collective nuclei. However, this rule has been challenged by recent measurements on some neutron-deficient nuclei
near $N_\mathrm{n}=90$ and the proton dripline~\cite{Grahn2016,Saygi2017,Cederwall2018,Goasduff2019,Zhang2021}, suggesting an anomalous collective motion characterized by $R_{4/2}>2.0$ and $B_{4/2}\ll1.0$. Such an exotic phenomenon has not been observed in conventional modes nor be produced by calculations using microscopic approaches, thereby posing a significant challenge to nuclear models~\cite{Grahn2016}.

Recent studies~\cite{Zhang2022,Wang2020,Zhang2024,Pan2024,Teng2025,Zhang2025} based on the phenomenological version of IBM (called IBM-1) have suggested that the $B_{4/2}<1.0$ feature~\cite{Cakirli2004} may be
attributed to band-mixing effects. These effects can be modeled using a collective Hamiltonian that incorporates triaxial rotor modes~\cite{Leschber1987,Castanos1988,Smirnov2000,Zhang2014,Teng2024}, which mix or compete with other collective modes~\cite{Zhang2024,Teng2025}. Theoretically, the inclusion of triaxial rotor modes implies a connection to a certain degree of triaxial deformation~\cite{Bohrbook}, whether it manifests as an intrinsic deformation at the mean-field level or as a dynamic deformation at the quantal level~\cite{Zhang2025}. Here, dynamic triaxial deformation, also referred to as effective triaxial deformation~\cite{Castanos1984,Elliott1986}, signifies that the system exhibits large $\gamma$ fluctuation in its low-lying states. Typically, a sophisticated polynomial combination of various high-order terms~\cite{Heyde1984,Berghe1985,Vanthournout1988,Vanthournout1990,Thiamova2010,Rowe2005,Sorgunlu2008,Wang2023} is required to integrate triaxial rotor-like structures into the IBM-1 Hamiltonian~\cite{Smirnov2000,Zhang2014,Teng2024,Thiamova2010}, which to some extent obscure the inherent correlation between triaxial deformation and the band-mixing induced $B(E2)$ anomaly features in the IBM framework~\cite{Zhang2025}. In contrast, triaxial deformation can be naturally realized in the microscopic version of the IBM, known as the proton-neutron interaction model (IBM-2)~\cite{Iachellobook}, even within the SU(3) dynamical symmetry~\cite{Dieperink1982}. This provides a unique opportunity for analytical study of the potential correlation between
the $B(E2)$ anomaly phenomenon and triaxial deformation in the IBM systems~\cite{Zhang2025}. It should be noted that the IBM-2, which describes a two-fluid boson system distinguishing protons from neutrons~\cite{Arias2004,Caprio2004,Caprio2005}, has a robust shell-model basis~\cite{Iachello1987} compared to the IBM-1. Its microscopic aspects have been increasingly emphasized through a self-consistent derivation of the Hamiltonian based on the microscopic mean-field calculations~\cite{Nomura2008,Nomura2010,Nomura2011I}.

The primary objective of this study is to
explore whether the anomalous collective behavior, distinguished by $B_{4/2}<1.0$ and $R_{4/2}>2$, can be produced in a straightforward manner from the SU(3) symmetry within the two-fluid boson system as described by the IBM-2 framework~\cite{Iachellobook}.

\begin{center}
\vskip.2cm\textbf{II. Model Hamiltonian}
\end{center}\vskip.2cm

In the IBM~\cite{Iachellobook}, the building blocs are two types of boson (operators): the monopole $s$ boson with $J^\pi=0^+$ and the quadrupole $d$ boson with $J^\pi=2^+$. In the IBM-2, the degrees of freedom of protons are further
distinguished from those of neutrons, meaning that proton and neutron bosons are treated as distinct constituents. Consequently,
the IBM-2 models a system comprised of two interacting fluids~\cite{Caprio2005}, in contrast to the IBM-1, which does not differentiate between protons and neutrons and thus represents a single-fluid boson system.

For both proton or neutron bosons, the bilinear products of creation and
annihilation operators
\begin{equation}
G_{\alpha,\beta}^{(\rho)}=b_{\rho,\alpha}^\dag b_{\rho,\beta},~~{(\alpha,\beta=1,\cdots,6)}\
\end{equation}
comprise a total of 36 independent elements, generating the dynamical symmetry U(6)$_\rho$ with $\rho=\pi,\nu$. As a result, the IBM-2 possesses the direct product dynamical symmetry $\mathrm{U}_\pi(6)\otimes \mathrm{U}_\nu(6)$, which can be
reduced to the angular momentum symmetry group $\mathrm{SO}_{\pi+\nu}(3)$ through various group chains corresponding to different dynamical symmetry limits. Hamiltonian and physical operators in the IBM-2~\cite{Iachellobook} are constructed from these boson operators.
For example, the angular momentum operator and quadrupole moment operator for proton or neutron bosons are defined by
\begin{eqnarray}\label{QLI}
&&\hat{L}_{\rho}=\sqrt{10}(d_\rho^\dag\times\tilde{d}_\rho)^{(1)}\, ,\\ \label{QLII}
&&\hat{Q}_{\rho}^{\chi_\rho}=(d_\rho^\dag\times\tilde{s}_\rho+s_\rho^\dag\times\tilde{d}_\rho)^{(2)}+\chi_\rho(d_\rho^\dag\times\tilde{d}_\rho)^{(2)}\,
\end{eqnarray}
with the parameter $\chi_\rho\in[-\frac{\sqrt{7}}{2},~\frac{\sqrt{7}}{2}]$. Accordingly, the total angular momentum and quadrupole moment operators in the two-fluids systems are constructed as
\begin{eqnarray}\label{LQ}
&&\hat{L}=\hat{L}_\pi+\hat{L}_\nu,~~\hat{Q}=\hat{Q}_\pi^{\chi_\pi}+\hat{Q}_\nu^{\chi_\nu}\, .\end{eqnarray}
With these definitions, the SU(3) symmetry can be generated by the operators
$\hat{L}$ and $\hat{Q}$ under two parametrization schemes: $\chi_\pi=\chi_\nu=\pm\sqrt{7}/2$ and $\chi_\pi=-\chi_\nu=\pm\sqrt{7}/2$. The former parametrization can be also applied to the IBM-1 without distinguishing protons ($\pi$) and neutrons ($\nu$), yielding the SU(3) (prolate) limit with $\chi=-\sqrt{7}/2$ and the $\overline{\mathrm{SU(3})}$ (oblate) limit with $\chi=\sqrt{7}/2$, respectively.
In contrast, the SU(3) symmetry special to the two-fluid IBM-2 system is that realized by the latter parametrization (denoted by SU$^{\ast}$(3)), where the equilibrium configuration consists of a proton fluid with axially symmetric oblate (prolate) deformation coupled to a neutron fluid with axially symmetric prolate (oblate) deformation~\cite{Caprio2005}, resulting in an overall nuclear shape with triaxial deformation~\cite{Dieperink1982}.

To provide an analytical study of triaxial dynamics in the SU$^{\ast}$(3) limit, we consider a quadrupole-type Hamiltonian
\begin{eqnarray}\label{H}
\hat{H}=\kappa\hat{Q}\cdot\hat{Q}+\eta(\hat{L}\times\hat{Q}\times\hat{L})^{(0)}\,
\end{eqnarray}
with the parameters $\kappa<0$ and $\chi_\pi=-\chi_\nu=\chi=\pm\frac{\sqrt{7}}{2}$.
In contrast to that traditionally used~\cite{Dieperink1982,Caprio2004}, the present Hamiltonian includes a three-body term to yield band mixing effects when the two-fluid system falls into triaxial deformation. For convenience, we take $\chi=\frac{\sqrt{7}}{2}$ in the following discussions, implying that a oblate proton fluid coupled to a prolate neutron fluid
is assumed in this case~\cite{Caprio2005}. Then, the corresponding SU$^{\ast}$(3) dynamical symmetry is characterized by the group chain~\cite{Iachellobook},
\begin{eqnarray}\label{group}
&&\mathrm{U}_\pi(6)\otimes\mathrm{U}_\nu(6)\supset\overline{\mathrm{SU_\pi(3)}}\otimes \mathrm{SU}_\nu(3)\\ \nonumber
&&~~~~~~~~~~~~~~~~~~~~~~~~\supset \mathrm{SU}_{\pi\nu}^\ast(3)\supset\mathrm{SO}_{\pi\nu}(3)\supset\mathrm{SO}_{\pi\nu}(2)\, .
\end{eqnarray}
The eigenvectors of the Hamiltonian can be symbolized using the quantum numbers labeling the irreducible representation (IRREP) for each subgroup contained in the group chain. Given that the SU(3) symmetry-conserving three-body term in Hamiltonian (\ref{H})
can induce band mixing only at the level of $\mathrm{SU}_{\pi\nu}^\ast(3)\supset\mathrm{SO}_{\pi\nu}(3)$,
the eigenvector can be expanded in terms of the undisturbed SU(3) basis~\cite{Kotabook} as
\begin{eqnarray}\label{wave}
|\Psi\rangle_{\xi LM}=\sum_{\tilde{\chi}}C_{\tilde{\chi}}^\xi|\Phi\rangle_{\tilde{\chi} LM}\, ,
\end{eqnarray}
where $C_{\tilde{\chi}}^\xi$ are the expansion coefficients with $\xi$ representing all quantum numbers other than $L,~M$.
Specifically, the SU(3) basis in the IBM-2~\cite{Iachellobook} can be expressed as
\begin{eqnarray}\label{basis}
|\Phi\rangle_{\tilde{\chi} LM}=|N_\pi,N_\nu, (\lambda_\pi,\mu_\pi),(\lambda_\nu,\mu_\nu),\tilde{\chi}_s,(\lambda,\mu),\tilde{\chi},L,M\rangle\, ,
\end{eqnarray}
where $N_\rho$ is the number of proton or neutron bosons, $(\lambda_\rho,\mu_\rho)$ are the quantum numbers labeling the corresponding SU(3) group, and $\tilde{\chi}_s$ and $\tilde{\chi}$
denote additional quantum numbers in the SU(3) IRREPs reduction~\cite{Iachellobook}.
Consequently, the quantum numbers $\Big[N_\pi,N_\nu, (\lambda_\pi,\mu_\pi),(\lambda_\nu,\mu_\nu),\tilde{\chi}_s,(\lambda,\mu),\tilde{\chi},L,M\Big]$ all remain good ones in the SU(3)$^{\ast}$ symmetry limit after introducing band mixing, as indicated by Eq.~(\ref{wave}).
The labels used here are consistent with those adopted in \cite{Iachellobook}, except for the abbreviation $\tilde{\chi}_s\equiv (\tilde{\chi}_1,\tilde{\chi}_2,\tilde{\chi}_3)$.
If $(\lambda,\mu)$ are known, the allowed angular momentum quantum number can be extracted from
the rules~\cite{Kotabook}:
\begin{eqnarray}\label{L}\nonumber
&&K=\mathrm{min}[\lambda,\mu],~\mathrm{min}[\lambda,\mu]-2,~\mathrm{min}[\lambda,\mu]-4,\cdots,~0~\mathrm{or}~1\\
&&L=0,~2,~4,\cdots, ~\mathrm{max}[\lambda,\mu], ~\mathrm{for}~K=0,\\ \nonumber
&&L=K,~K+1,~K+2,\cdots, ~K+\mathrm{max}[\lambda,\mu],~\mathrm{for}~K>0,\\ \nonumber
&&M=-L,~-L+1,~-L+2,~\cdots,~L\, ,
\end{eqnarray}
in which $\mathrm{min}[a,b]$ ($\mathrm{max}[a,b]$) denotes the minimal (maximum) value between $a$ and $b$.

The eigenvalues solved from the equation $\hat{H}|\Psi\rangle=E|\Psi\rangle$ can be expressed by
\begin{eqnarray}\label{E}
E&=&\frac{\kappa}{2}\Big[(\lambda^2+\mu^2+\lambda\mu+3\lambda+3\mu)-\frac{3}{4}L(L+1)\Big]\\ \nonumber
~~~~&+&\eta\langle(\hat{L}\times\hat{Q}\times\hat{L})^{(0)}\rangle\, ,
\end{eqnarray}
where the last term represents the expectation value of the three-body interaction within the eigenvectors given in
(\ref{wave}). In the SU$^{\ast}$(3) limit, the SU(3) IRREP for the ground state are determined by
$(\lambda,\mu)=(2N_\nu,2N_\pi)$, where $N_\pi$ ($N_\nu$) represents the number of valence proton (neutron) pairs counted from the nearest closed shell.
The $\gamma$ deformation of the SU(3) system can be estimated using the formula
\begin{equation}\label{su3gamma}
\gamma=\mathrm{tan}^{-1}\Big(\frac{\sqrt{3}(\mu+1)}{2\lambda+\mu+3}\Big)\,
\end{equation}
proposed in \cite{Castanos1988}, which has been adopted in the IBM-1 to indicate the $\gamma$ deformation in the SU(3) mapping of triaxial rotor~\cite{Zhang2022,Zhang2024,Smirnov2000,Zhang2014}.
This simple formula allows us to deduce directly that axially asymmetric deformation will be exhibited in the SU$^{\ast}$(3) symmetry in the two-fluid IBM-2 system. For example, the maximally triaxial deformation with $\gamma=30^\circ$ is achieved when $N_\nu=N_\pi$, which is consistent the earlier mean-field analysis of this symmetry limit~\cite{Dieperink1982,Caprio2005}.
Other configurations can be obtained through the tensor products of SU(3) IRREPs, $(\lambda_\pi,\mu_\pi)\otimes(\lambda_\nu,\mu_\nu)=\oplus\sum(\lambda,\mu)$, with~\cite{Dieperink1982}
\begin{eqnarray}
&(\lambda_\pi,\mu_\pi)=(0,2N_\pi),~(2,2N_\pi-4),~(4,2N_\pi-8)~\cdots,\\
&(\lambda_\nu,\mu_\nu)=(2N_\nu,0),~(2N_\nu-4,2),~(2N_\nu-8,4)~\cdots\, ,
\end{eqnarray}
yielding
\begin{eqnarray}\nonumber
(\lambda,\mu)&=&(2N_\nu,2N_\pi),~(2N_\nu-4,2N_\pi+2),\\
&~&(2N_\nu+2,2N_\pi-4),~\cdots\, .
\end{eqnarray}
Using these rules for quantum numbers, one can analytically evaluate the excitation energies for specific cases.
In general, the last term, $\langle(\hat{L}\times\hat{Q}\times\hat{L})^{(0)}\rangle\equiv\langle LQL\rangle$, in Eq.~(\ref{E}) cannot be analytically derived except for the cases of $L=0$, where this term vanishes. In other words, $E(0_n^+)$ remains unaffected by the three-body term. Another important analytically solvable case is the one associated with the $(\lambda,\mu=2)$ configuration, which is permissible
for the ground-state IRREP where both $\lambda$ and $\mu$ are even numbers~\cite{Iachellobook}.

In the subsequent discussion, we focus on analyzing states built upon the ground-state SU(3) configuration, which is pertinent to the $B(E2)$ anomaly. The analysis can be readily extended to excited SU(3) IRREPs, where $(\lambda,\mu)$ are not restricted to even numbers.
For the case of $(\lambda,\mu=2)$, one can derive the following analytical expressions:
\begin{eqnarray}\label{E21}
&\langle LQL\rangle_{2_1^+}=-\sqrt{\frac{3(4\lambda^2+20\lambda+49)}{20}},\\
&\langle LQL\rangle_{2_2^+}=\sqrt{\frac{3(4\lambda^2+20\lambda+49)}{20}},\\
&\langle LQL\rangle_{4_1^+}=-\frac{7\sqrt{15}(5+2\lambda)}{30}-\frac{\sqrt{(60\lambda^2+300\lambda+5775)}}{10},\\
&\langle LQL\rangle_{4_2^+}=-\frac{7\sqrt{15}(5+2\lambda)}{30}+\frac{\sqrt{(60\lambda^2+300\lambda+5775)}}{10},\\ \label{E51}
&\langle LQL\rangle_{3_1^+}=0,~~~~\langle LQL\rangle_{5_1^+}=-\frac{3\sqrt{15}(2\lambda+5)}{10},\\
&\langle LQL\rangle_{6_1^+}=-\frac{3\sqrt{15}(5+2\lambda)}{5}-\frac{\sqrt{5(12\lambda^2+60\lambda+5115)}}{10},\\
&\langle LQL\rangle_{6_2^+}=-\frac{3\sqrt{15}(5+2\lambda)}{5}+\frac{\sqrt{5(12\lambda^2+60\lambda+5115)}}{10},\\
&\langle LQL\rangle_{7_1^+}=-\frac{11\sqrt{15}(5+2\lambda)}{15},~\langle LQL\rangle_{9_1^+}=-\frac{13\sqrt{15}(5+2\lambda)}{10},\\ \nonumber
&\cdots\, .
\end{eqnarray}
Clearly, the $(\lambda,\mu=2)$ configuration with $\lambda\geq2$ yields a two-band system. This includes the ground-state band with $L^\pi=0_g^+~,2_g^+,~4_g^+,\cdots$ up to $L_g=\lambda$, and the $\gamma$-band with $L^\pi=2_\gamma^+~,3_\gamma^+,~4_\gamma^+,\cdots$ up to $L_\gamma=\lambda+2$. The results indicate that the three-body term in the Hamiltonian (\ref{H}) with $\eta>0$ lowers the level energies of the ground-state band but increase those of the $\gamma$ band with $L_\gamma=$even. If removing this three-body term, level energies of both the ground-state band and $\gamma$ band may follow the $L(L+1)$ rule, as indicated by Eq.~(\ref{E}).
The $B(E2)$ transitions can be evaluated via
\begin{eqnarray}
B(E2;L_i\rightarrow L_f)=\frac{|\langle\xi_fL_f\parallel T(E2)\parallel\xi_iL_i\rangle|^2}{2L_i+1}\,
\end{eqnarray}
with the $E2$ operator defined as \begin{eqnarray}\label{E2}
T(E2)=e_\mathrm{B}\hat{Q}\, ,\end{eqnarray}
where $e_\mathrm{B}$ represents the effective charge and the $\hat{Q}$ operator is taken as same as that involved in the Hamiltonian (\ref{H}).
For simplicity, we have assumed here that the boson effective charges for protons and neutron take the same values~\cite{Nomura2010}, which means that the $B(E2)$ transition are only allowed between states within the same SU(3) IRREP. In principle, similarly to the excitation energies, wave functions ($C_{\tilde{\chi}}^\xi$) for a two-band system and thus the $B(E2)$ results can also be expressed analytically as functions of $\lambda$, using the SU(3) Wigner coefficients tabulated in \cite{Vergados1968}. However, the resulting expressions are too lengthy to provide meaningful reference. So, only analytical expressions for excitation energies in the cases  of $(\lambda,\mu=2)$ are provided as shown above.

\begin{center}
\vskip.2cm\textbf{III. Results and Discussions}
\end{center}\vskip.2cm

\begin{figure}
\begin{center}
\includegraphics[scale=0.3]{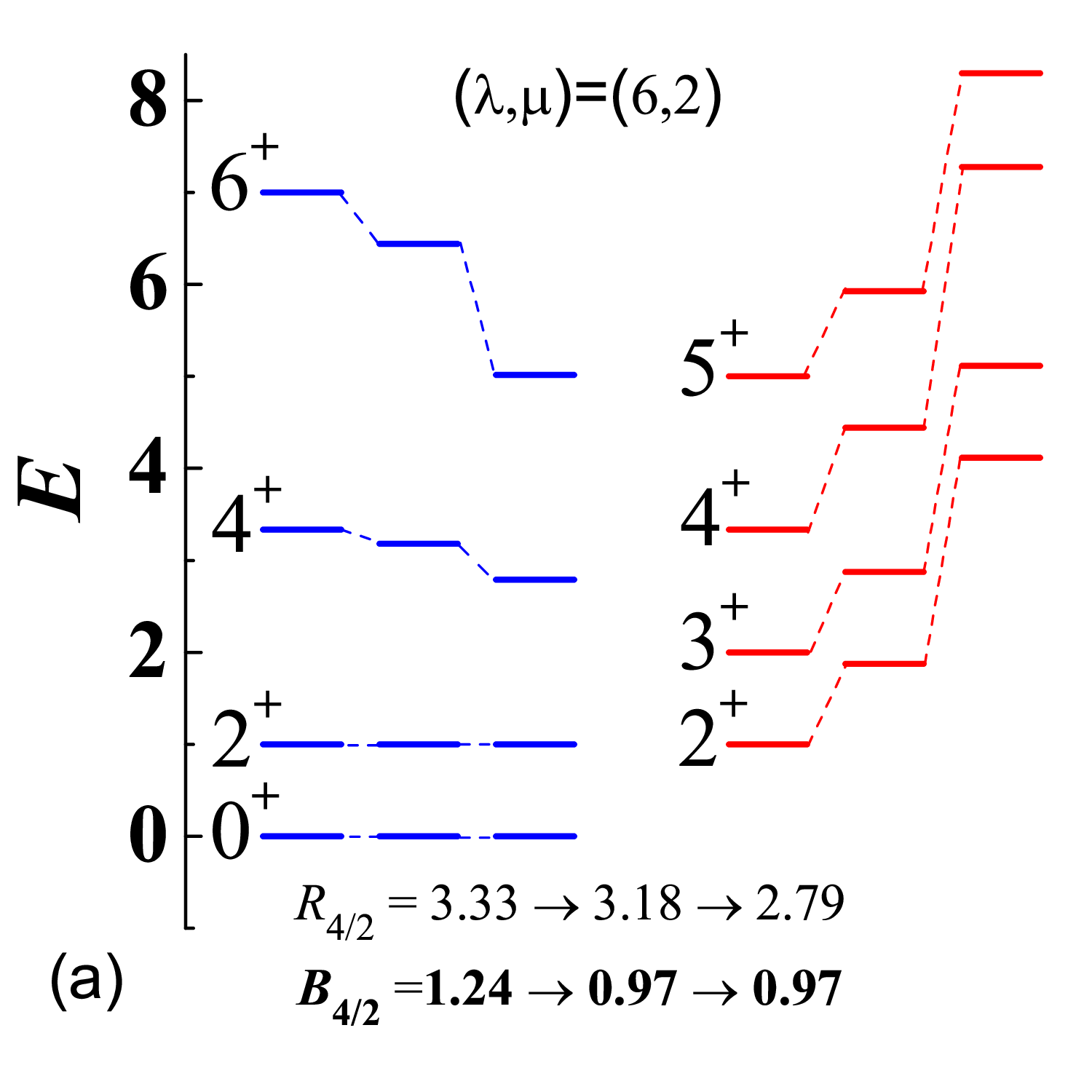}
\includegraphics[scale=0.3]{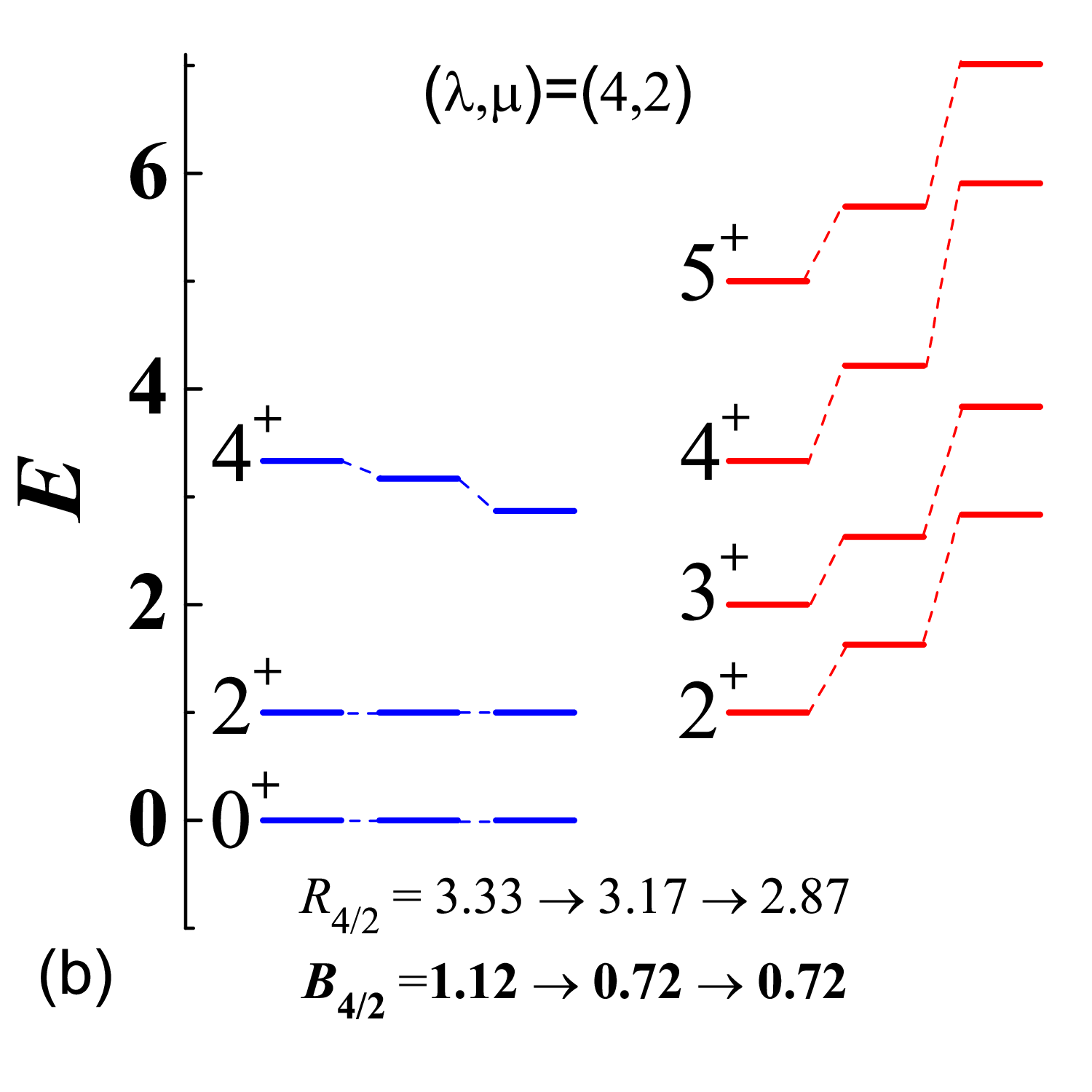}
\caption{(a) The low-lying level patterns for $(\lambda,\mu)=(6,2)$, obtained for $\eta/\kappa=0.0,-0.1,-0.2$ (from left to right), are shown. (b) The same as in (a) but for $(\lambda,\mu)=(4,2)$. In each panel, the level energies have been normalized to $E(2_1^+)=1.0$. }\label{F1}
\end{center}
\end{figure}

To illustrate band mixing in a two-band SU$^{\ast}$(3) system, the level patterns for $(\lambda,\mu)=(6,2)$ and $(\lambda,\mu)=(4,2)$, obtained from solving the Hamiltonian (\ref{H}), are presented in Fig.~\ref{F1}.
As observed in Fig.~\ref{F1}(a), significant changes in the normalized level energies occur as $|\eta/\kappa|$ increases, which aligns with the analytical expressions given in (\ref{E21})-(\ref{E51}). More importantly, the results demonstrate that the three-body term in the SU$^{\ast}$(3) limit can induce a $B_{4/2}<1.0$ feature while preserving the regularity of level pattern.
In contrast to $R_{4/2}$, it is evident that the $B_{4/2}$ value remains independent of the absolute values of $\eta/\kappa$ when $\eta\neq0$. This suggests that $B_{4/2}<1.0$ phenomenon caused by band mixing is an intrinsic characteristic of the SU$^{\ast}$(3) limit. This observation can be understood by noting that the wave functions for a given $(\lambda,\mu)$ and $L$ are primarily determined by the presence of the three-body term, as indicated by (\ref{E}). Consequently, the strength parameter $\eta$ acts as a scaling factor for all states within a given $(\lambda,\mu)$, leaving the wave functions unchanged with respect to $\eta$.

As indicated by the IBM-1 analysis~\cite{Zhang2022,Zhang2024,Zhang2025}, the anomalous $E2$ behavior of yrast states tends to in occur in systems characterized by intrinsic or dynamic triaxial deformation. In the SU$^{\ast}$(3) limit, $\gamma$ deformations can be estimated from Eq.~(\ref{su3gamma}), yielding $\gamma\simeq17^\circ$ for $(\lambda,\mu)=(6,2)$. In addition to $B_{4/2}=0.97$, the $(\lambda,\mu)=(6,2)$ configuration results in $B(E2;6_1^+\rightarrow4_1^+)/B(E2;2_1^+\rightarrow0_1^+)=0.61$ and $B(E2;8_1^+\rightarrow6_1^+)/B(E2;2_1^+\rightarrow0_1^+)=0.0$. From these results, two conclusions can be drawn: (a)~$B(E2)$ anomaly extends to yrast states with higher spins; (b)~Transition $B(E2;L_1^+\rightarrow(L-2)_1^+)$ for $L>\lambda$ is forbidden due to the termination of the ground-state band at $L=\mathrm{max}[\lambda,\mu]$. These findings serve as signatures for a two-fluid system dominated by the SU$^{\ast}$(3) symmetry with the ground-state configuration $(\lambda,\mu)=(2N_\pi,2N_\nu)$ or $(2N_\nu,2N_\pi)$. As shown in Fig.~\ref{F1}(b), the two-band system for $(\lambda,\mu)=(4,2)$ exhibits dynamic features similar to those of $(\lambda,\mu)=(6,2)$. However, the $B_{4/2}$ ratio for $(\lambda,\mu)=(4,2)$, corresponding to $\gamma\simeq22^\circ$, is notably lower than that for $(\lambda,\mu)=(6,2)$, which corresponds to $\gamma\simeq17^\circ$. This highlights the influence of triaxial deformation on the $B(E2)$ structure.

\begin{figure}
\begin{center}
\includegraphics[scale=0.3]{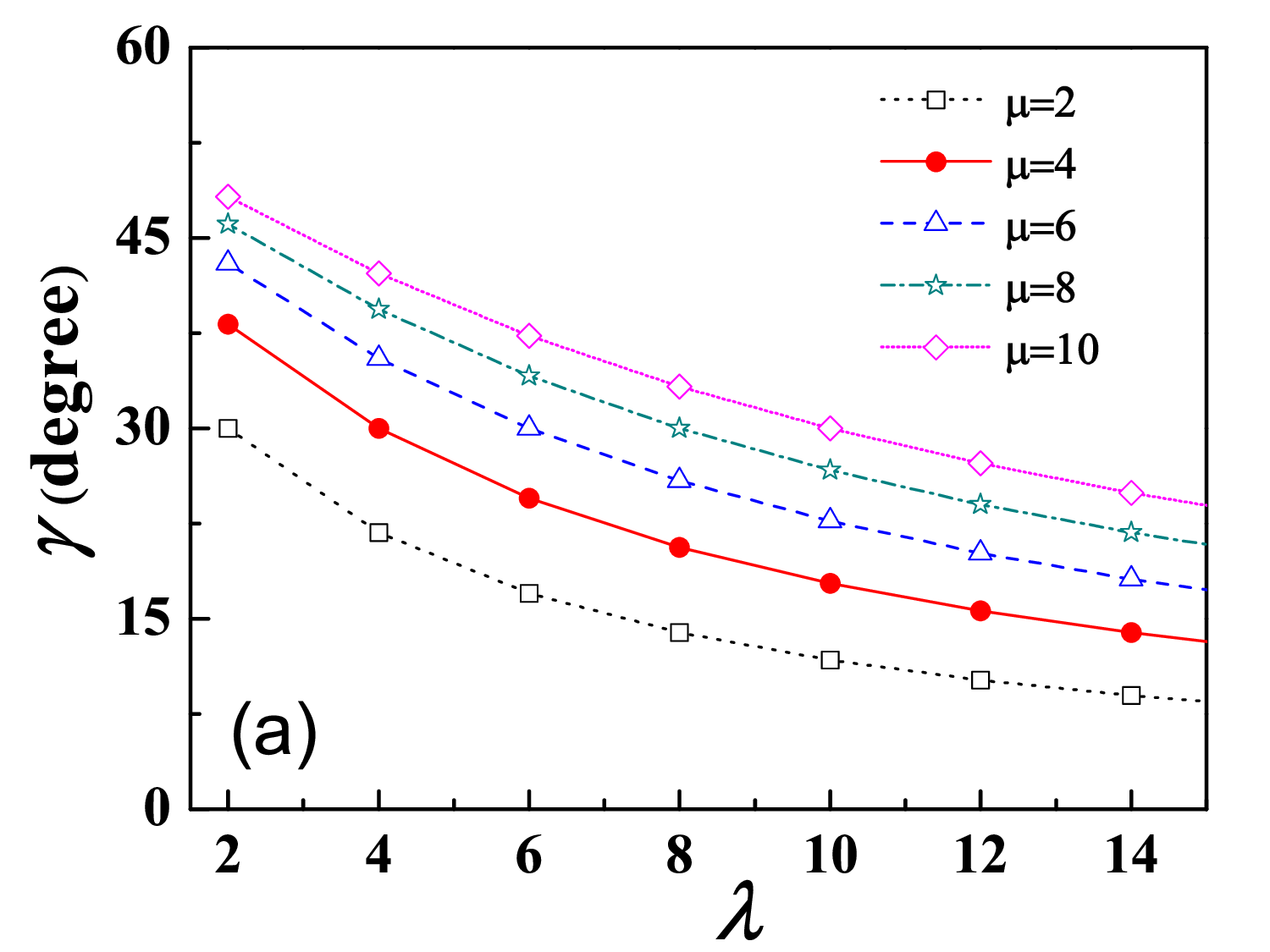}
\includegraphics[scale=0.3]{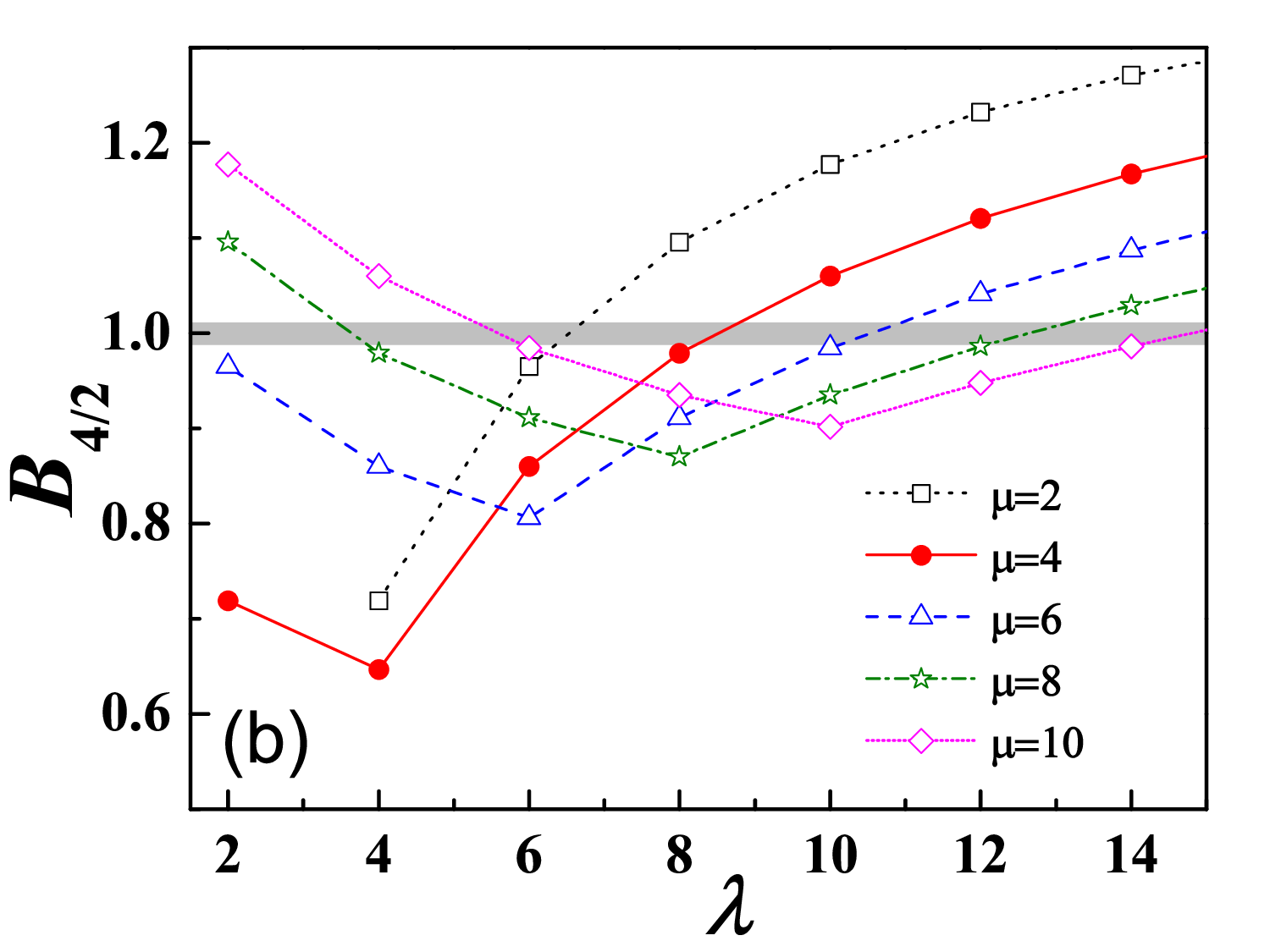}
\caption{(a) The $\gamma$ deformations for different ($\lambda,\mu$), calculated based on Eq.~(\ref{su3gamma}), are presented as a function of $\lambda$. (b) The corresponding $B_{4/2}$ ratio for different ($\lambda,\mu$) solved from the Hamiltonian (\ref{H}) are shown as a function of $\lambda$.  For $\lambda\geq\mu$, the parameter in calculation is set to $\eta/\kappa=-0.05$, while for $\lambda<\mu$, it is set to $\eta/\kappa=0.05$. In addition, $B_{4/2}$ for $(\lambda,\mu)=(2,2)$ is not shown because the ground-state band in this case terminates at $2_1^+$.  }\label{F2}
\end{center}
\end{figure}

The analysis for $(\lambda,\mu)=(\lambda,2)$ can be directly extended to more general SU(3) IRREPs. Nevertheless, for cases with $(\lambda,\mu>2)$, model solutions can only be obtained numerically. Due to the symmetric properties of the SU(3) Wigner coefficients under the transformation $(\lambda,\mu)\rightarrow(\mu,\lambda)$~\cite{Vergados1968}, the energies and $B(E2)$ transitions derived from $(\lambda,\mu)=(a,b)$ are identical to those from $(\lambda,\mu)=(b,a)$ after changing the parameter $\eta$ in (\ref{H}) to $-\eta$. This implies that $\eta<0$ must be used for cases where $\lambda<\mu$ to ensure the regularity of yrast levels expected for a collective mode.
In addition, it can be verified that interchanging $\lambda$ and $\mu$ transforms the $\gamma$ deformation from $\gamma=A$ to $\gamma=60^\circ-A$, as well as the quadrupole moment from $\hat{Q}(2_1^+)=B$ to $\hat{Q}(2_1^+)=-B$. To systematically analyze the SU$^{\ast}$(3) system, the $\gamma$ deformation and $B_{4/2}$ ratio obtained from different $\mu$ values are plotted as functions of $\lambda$ in Fig.~\ref{F2}. As shown in Fig.~\ref{F2}(a), the $\gamma$ values decrease monotonically with increasing $\lambda$ within the range $\gamma\in(0^\circ,60^\circ)$, reaching $\gamma=30^\circ$ at $\lambda=\mu$, which corresponds to maximal triaxial deformation. In contrast, Fig.~\ref{F2}(b) illustrates that the $B_{4/2}$ ratio as a function of $\lambda$ exhibits non-monotonic behavior, with minima occurring at $\lambda=\mu$. Consequently, the lowest $B_{4/2}$ ratios, potentially leading to $B_{4/2}<1.0$, favor maximal triaxialilty. It is important to note that the $B_{4/2}$ results presented in Fig.~\ref{F2}(b) are independent of the absolute value $|\eta/\kappa|$ for each given $(\lambda,\mu)$, as confirmed by numerical calculations. Therefore, the observed $B(E2)$ anomaly in the SU$^{\ast}$(3) limit is indeed an intrinsic feature resulting from band mixing.

\begin{figure}
\begin{center}
\includegraphics[scale=0.35]{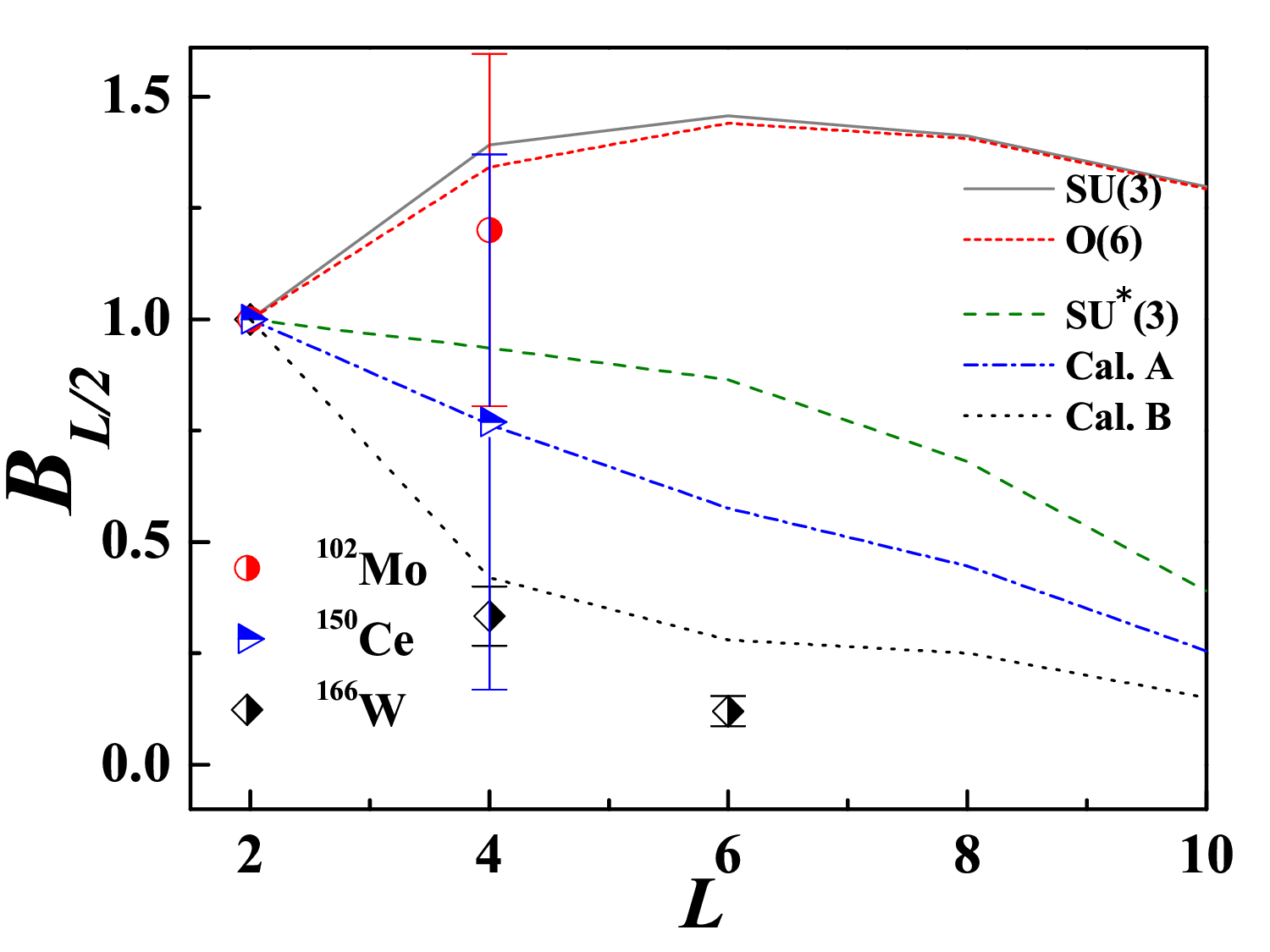}
\caption{The cascades of $B_{L/2}\equiv B(E2;L_1^+\rightarrow(L-2)_1^+)/B(E2;2_1^+\rightarrow0_1^+)$ obtained for the SU(3) limit, O(6) limit, SU$^{\ast}$(3) limi, as well as two examples (labeled as Cal.A and Cal.B) deviating from the SU$^{\ast}$(3) limit are presented (the parameters are detailed in the text). Experimental data~\cite{Saygi2017,Frenne2009,Basu2013} for nuclei corresponding to $(N_\pi,~N_\nu)=(4,~5)$ are included for comparative analysis.}\label{F3}
\end{center}
\end{figure}

To gain a deeper understanding of the $B(E2)$ structure of yrast band, Fig.~\ref{F3} presents the cascades defined as $B_{L/2}\equiv B(E2;L_1^+\rightarrow(L-2)_1^+)/B(E2;2_1^+\rightarrow0_1^+)$, calculated under various symmetry limits with $(N_\pi,N_\nu)=(4,5)$ and two cases breaking the ideal SU$^{\ast}$(3) dynamical symmetry. Additionally, experimental data for $^{102}$Mo~\cite{Frenne2009}, $^{150}$Ce~\cite{Basu2013} and $^{166}$W~\cite{Saygi2017}, all characterized by $(N_\pi,~N_\nu)=(4,~5)$, are also included for comparative analysis. These experimental data are incorporated here to illustrate typical realistic scenarios rather than to achieve the best fit in theory. In the calculations, the Hamiltonian parameters are set as follows: $(\chi_\pi,~\chi_\nu,~\eta/\kappa)=(-\sqrt{7}/2,~-\sqrt{7}/2,~-0.1)$ for the SU(3) limit, $(\chi_\pi,~\chi_\nu,~\eta/\kappa)=(0,~0,~-0.1)$ for the O(6) limit, and $(\chi_\pi,~\chi_\nu,~\eta/\kappa)=(\sqrt{7}/2,~-\sqrt{7}/2,~-0.1)$ for the SU$^{\ast}$(3) limit. For scenarios deviating from the SU$^{\ast}$(3) limit, referred to as SU$^{\ast}$(3) breaking, we present two examples where $\chi_\pi=\sqrt{7}/2$ but $\chi_\nu\neq-\sqrt{7}/2$. Specifically, the parameter sets are chosen as $(\chi_\pi,~\chi_\nu,~\eta/\kappa)=(\sqrt{7}/2,~-0.7,~-0.15)$ and $(\sqrt{7}/2,~-0.3,~-0.26)$, respectively. The corresponding results, labeled as Cal.A and Cal.B, are illustrated in Fig.~\ref{F3}.
To perform calculations in cases beyond the SU(3) symmetry limit, we have adopted the numerical scheme proposed in \cite{Hu2023} for solving the IBM-2 Hamiltonian.
As shown in Fig.~\ref{F3}, the ratio values in the SU(3) and O(6) limits consistently exhibit $B_{L/2}\geq1.0$, even when incorporating the three-body term as described in Eq.~(\ref{H}). Conversely, a monotonic decrease in the cascades as a function of $L$ is observed in the SU$^{\ast}$(3) limit, which generates the ground-state IRREP $(\lambda,\mu)=(10,8)$ for $(N_\pi,~N_\nu)=(4,~5)$.
This indicates that the $B(E2)$ anomaly in the SU$^{\ast}$(3) limit, characterized by $B_{L/2}<1.0$, persists throughout the entire yrast line, as exemplified by the $(\lambda,\mu=2)$ cases discussed earlier. Furthermore, such anomalous features are accentuated
in the SU$^\ast$(3) breaking scenarios, which appears to be supported by the available data for $^{150}$Ce and $^{166}$W.
In contrast, the data for $^{102}$Mo align more closely with the "normal" modes described by the SU(3) or O(6) limits. These findings, to some extent, provide evidences for the band-mixing induced $B(E2)$ anomaly in the IBM-2 near the triaxial SU$^{\ast}$(3) limit.

Although the mean-field calculations suggest~\cite{Caprio2005} that the ideal SU$^{\ast}$(3) system in the large-$N$ limit comprise a proton fluid with oblate (prolate) deformation coupled to a neutron fluid with prolate (oblate) deformation~\cite{Caprio2005}, this does not imply that a realistic scenario with opposite sign of $\chi_\pi$ and $\chi_\nu$ is necessarily associated with completely inverse $\gamma$ deformations between protons and neutrons. This point can be partially understood by examining the calculated quadrupole moments for proton and neutron bosons via $\hat{Q}_\rho(L^+)=\sqrt{\frac{16\pi}{5}}\langle LM|\hat{Q}_\rho|LM\rangle_{M=L}$. For instance, adopting the same parameters as those used in Fig.~\ref{F3}, it is shown both $\hat{Q}_\pi(2_1^+)<0$ and $\hat{Q}_\nu(2_1^+)<0$ for the case of Cal.A, while both $\hat{Q}_\pi(2_1^+)>0$ and $\hat{Q}_\nu(2_1^+)>0$ for the case of Cal. B. This suggests that the $\gamma$ deformations of low-spin states are influenced not only by the $\chi_{\pi(\nu)}$ values but also by the boson numbers $N_{\pi(\nu)}$ and the strength ratio $\eta/\kappa$, when using the simple Hamiltonian form (\ref{H}). Additionally, the results indicate that relatively large fluctuations in $\gamma$ are anticipated for the $B(E2)$ anomaly systems when deviating from the ideal SU$^{\ast}$(3) limit. This observation aligns with the recent analysis of $B(E2)$ anomaly within the IBM-1 framework presented in \cite{Zhang2025}. That study demonstrates that the $B(E2)$ anomaly features arising from band mixing can be modeled even using a Hamiltonian with zero $\gamma$ deformation at the mean-field level, yet exhibiting pronounced $\gamma$ softness. As a result, a meaningful evaluation of the current theoretical perspective involves finding a method to determine the effective $\gamma$ deformation~\cite{Werner2005} or $\gamma$ fluctuations in $B(E2)$ anomaly systems. Further discussions on $\gamma$ fluctuations within the IBM-2 framework will be addressed elsewhere.

\begin{center}
\vskip.2cm\textbf{IV. Summary}
\end{center}\vskip.2cm

In summary, a theoretical analysis of the potential correlation between the $B(E2)$ anomaly phenomenon and triaxial deformation has been conducted within the framework of the IBM-2. The results demonstrate that the anomalous $E2$ behaviors, characterized by $B_{4/2}<1.0$, can naturally arise from the SU$^{\ast}$(3) dynamical symmetry limit of the IBM-2 when a three-body term is incorporated alongside the quadrupole-quadrupole interaction. This offers a clear symmetry-based interpretation of the $B(E2)$ anomaly features arising from band mixing, which were identified in the IBM-1 but described through complex polynomial combinations of various high-order terms~\cite{Zhang2022,Wang2020,Zhang2024,Pan2024,Teng2025,Zhang2025}. Beyond the analytic explanation based on the ideal SU$^{\ast}$(3) limit, which models a triaxially deformed two-fluid system, it is shown that the SU$^{\ast}$(3) breaking effects are crucial for providing a qualitative description of the experimentally observed $B(E2)$ anomaly. This conclusion is supported by available experimental data for relevant nuclei. In addition to even-even systems, similar observations of the $B(E2)$ anomaly phenomenon have also been reported in odd-A species~\cite{Zhang2021}. Extending this symmetry-based analysis to odd-A systems~\cite{Teng2025} and to other SU(3)-related microscopic framework~\cite{Bonatsos2017I,Bonatsos2017II} would be interesting. Related work is in progress.
\bigskip

\begin{acknowledgments}
Support from the National Natural Science Foundation of China under No.12375113 is acknowledged.
\end{acknowledgments}



\end{document}